\documentstyle[12pt,aasms4]{article}
\slugcomment{revised manuscript for AJ, submitted 23 July 96}
\lefthead{Zacharias, N. et al.}
\righthead{Twin Astrographic Catalog}
\begin{document}

\title{The Twin Astrographic Catalog (TAC) Version 1.0}
\author{N. Zacharias\altaffilmark{1}, M.I. Zacharias\altaffilmark{1},
       G.G. Douglass, G.L.Wycoff} 
\affil{U.S. Naval Observatory, 3450 Mass. Ave. N.W., Washington D.C. 20392,
       nz@pyxis.usno.navy.mil}

\altaffiltext{1}{with Universities Space Research Association (USRA),
  Division of Astronomy and Space Physics, Washington D.C.} 

\begin{abstract}
A first version of the Twin Astrographic Catalog (TAC) of positions 
for 705,679 stars within $-18^{\circ} \le \delta \le 90^{\circ}$ has been produced. 
The sky coverage of the TAC is complete to over 90\% in that area.
The limiting magnitude is about B=11.5.
Positions are based on $4912$ plates taken with the U.S. Naval Observatory
Twin Astrograph (blue, yellow lens) at epochs 1977--1986.
The TAC is supplemented by proper motions which are obtained from a combination with
a re--reduced Astrographic Catalog (AC). Some AC zones are available now and 
a complete northern hemisphere is expected by fall 1996.
Proper motions of almost all TAC stars will be generated as the AC work progresses.
The average precision of a catalog position is 90 mas per coordinate at epoch
of observation.
A large fraction of that error is introduced by the currently available 
reference stars. The inherent precision of the TAC data is considerably better.
The precision of the proper motions is currently 2.5 to 4 mas/yr.
Magnitude--dependent systematic errors have been found and preliminarily corrected. 
The final reduction of this plate material will be performed with the
Hipparcos catalog in 1997.
The TAC is about 3 times more precise than the PPM or ACRS in the northern
hemisphere at current epochs and contains about 3 times more stars.
The TAC has a higher star density than the Tycho catalog and provides independent,
high precision positions for a large fraction of the Tycho stars at an epoch 
about 10 years earlier than the Tycho mean epoch.
The TAC version 1.0 data are released as the AC zones become available.
For latest information, look at the US Naval Observatory
World Wide Web page http://aries.usno.navy.mil/ad/tac.html.

\end{abstract}


\section{INTRODUCTION}

The Twin Astrographic Catalog (TAC) contains positions and proper motions 
of stars to $B \approx 11.5$, with an average density
of $\approx 30$ stars per square degree on the 100 mas accuracy level
at current epochs. 
Its first application is to provide reference stars for
the Sloan Digital Sky Survey (SDSS) project (\cite{SDSS}).
The TAC covers the entire SDSS area north of $-18^{\circ}$ declination,
but not the complete northern hemisphere.
Observing is complete, however about 250 plates have not yet been measured.
The original plan for an all--sky survey (\cite{allsky}) has been altered 
and no plates have been taken south of $-18^{\circ}$.

There are 2 previous publications about the TAC data.
First, the Zodiacal Zone Catalog, based on part of the plate material 
used for the TAC, has been published (\cite{DH90})
in order to assist solar system observations.
These plates were remeasured to the TAC magnitude limit.
Second, preliminary results of the TAC reductions have been 
presented recently at an AAS meeting (\cite{ZD95}).

This paper presents the first version of the catalog
as reduced in the J2000/FK5 system.
A future version will be based on the Hipparcos catalog (\cite{Hip}),
which will become generally available in 1997.
Concurrent with the Hipparcos catalog is the Tycho catalog with
about a million stars over the entire sphere.
Most of the Tycho positions will be more accurate than the TAC positions,
but the TAC has a higher star density and the epoch difference
(TAC -- Tycho) of about 10 years is valuable for improving the proper
motions of stars common to both catalogs.
The TAC aids tremendously in the search for high proper motion stars.

The TAC is a large improvement over the AGK3 (\cite{AGK3}),
the previously best observed catalog in the northern hemisphere.
It will be comparable in precision to the CPC2 (\cite{CPC2I}, \cite{CPC2II})
for recent epochs on the southern hemisphere, but the TAC will be
more than twice as dense as the CPC2. 
In all areas of the sky which are covered by the TAC,
it will supersede the compiled catalogs PPM north (\cite{PPMN}),
PPM south (\cite{PPMS}) and the ACRS (\cite{ACRS}) for current epoch
applications.

The project started in the late 1970's at the US Naval Observatory,
using the Twin Astrograph equipped with 2 lenses which are corrected
for the blue and yellow spectral bandpasses.
Observations were made simultaneously with both lenses with a mean
epoch around 1982.
Proper motions have been obtained in combination with a complete
re--reduction of the Astrographic Catalog (AC) (\cite{urb95}, \cite{urb96}),
which provides the first epoch data.
Not all AC zones have been reduced yet.
The completion of the northern hemisphere AC zones is expected
in summer 1996.
Providing only positions without proper motions is not very helpful
for the general user.
Thus, we will make the TAC including proper motions available 
only for those areas where AC positions have been provided. 
Please follow the development on the World Wide Web pages of the US Naval Observatory,
http://aries.usno.navy.mil/ad/tac.html.
A CD-ROM version of the TAC is planned after the re--reduction with
the Hipparcos catalog is completed by the end of 1997.

\section{OBSERVATIONS}

Plates were taken with the US Naval Observatory 8--inch
Twin Astrograph (TA) simultaneously for the same field centers with 2 lenses.
The 2 lenses of 2--meter focal length are corrected for the blue and the yellow
spectral bandpass.
Table 1 summarizes details about the telescope.

The TA was guided in one axis (RA) only, with a photomultiplier
knife--edge autoguider using a 15 cm telescope mounted to the TA.
The project started with 10--minute exposure times for the blue and yellow plates.
Because of the bright Washington sky background, all plates taken since 1980.69 
have 4--minute exposure time.
All plates were taken between 1977.08 and 1985.71. 
Fig.~1 shows the distribution of plate epochs as a function of declination
of the plate centers.
For this version of the TAC, 2458 plates taken with the yellow lens
and 2454 plates taken with the blue lens have been used, out of a total
of $\approx 5180$  available plates.

A 2--fold center in corner overlap pattern per color was adopted with offsets 
of $2^{\circ} 15'$ in declination and $2^{\circ} 30'$ in right ascension between
adjacent field centers.
Thus, nominally each star appears on 4 plates.
Fig.~2 shows the sky distribution of the plates used for the catalog.
The gaps near the north celestial pole and in the most southern bands
will eventually be covered, once all available plates are measured.
A few single plates in the covered area are waiting for re--measurement
or are missing.
Plates with centers in the declination ranges
$-15^{\circ} \le \delta \le -9^{\circ}$ and
$+65^{\circ} \le \delta \le +90^{\circ}$ for the right ascension ranges 
$0^{h} \le \alpha \le 6^{h}$ and
$20^{h} \le \alpha \le 24^{h}$ are yet to be measured.

The TA plates were measured on the US Naval Observatory
STARSCAN measuring machine (\cite{DH90}) with a precision of
$ 0.5 $ to $ 0.7 \ \mu m$.
This error estimate was obtained from the comparison of direct and
reverse measurement of each plate and measurements of a calibrated grid plate.
Figure 3 shows the distribution of measurement date as a function of declination.
Most of the time the blue and yellow plate of each field was 
measured on the same day.

Although the STARSCAN measuring machine automatically determines each
star image center, most of the measures were obtained with an operator
manually supervising the data acquisition.
Some double stars could not be centered by the machine and manual hand
measures were made by the operator.
These measures turned out to be very inaccurate and all hand measures
were excluded from this version of the TAC.
Thus, some stars within the magnitude range $5^{m} \le B \le 11^{m}$
will not be in the TAC if the images appear blended on the plates.
All images within 5 arcseconds were assigned to the same star
and have a single entry in the catalog.
Stars brighter than about $5^{m}$ are not measured due to overexposure
problems.
After 1994 all plates were measured in a completely automatic mode.
A measured magnitude (opacity) was obtained for each image,
which serves as an instrumental magnitude and 
is approximately linear to the apparent brightness of that star.

A preliminary version of the Astrographic Catalogue (AC) served as input list
for the plate measuring process for the plates centered on declination 
$-6^{\circ} 45' \le \delta \le +29^{\circ} 15'$.
For the rest of the observing program, a preliminary version of the
Tycho Input Catalogue (TIC) was used.
A catalog magnitude was available for each star from its input list.
For the TIC stars a V magnitude was carried through the plate measuring
and reduction process, while most stars from the AC had a B magnitude.
Input lists with a magnitude cutoff of 11.0 was used for most of the plates.
Plates measured in 1995 used a cutoff of 12.0.

\section{PLATE REDUCTION}

\subsection{The Basics}

All TAC plates were reduced individually with
the International Reference Stars (IRS) catalog (\cite{co78}, 1995),
which is close to the J2000/FK5 system.
A minimum of 7, a maximum of 33 and an average of 21.5 reference stars
were found on individual plates.
Software from the Hamburg Block Adjustment Program Package (HBAPP) (\cite{HBAPP})
was used for all reductions.
Additional data checking and evaluation software was written
particularly for this project.
Rigorous apparent places and atmospheric refraction routines were 
applied to the reference stars' mean positions
and the weighted least--squares adjustment was performed 
with the observed places in the focal plane.
The random error for each star position in this adjustment step consists of 2 parts,
the nearly constant error from the plate measuring process,
assuming $\sigma_{xy} = 100$ mas,
and the individual reference star position error at the epoch of the plate.
Thus weights were obtained from the quadratic sum of both contributions.
The effective wavelength (midpoint of range as given in Table 1) was used
for the refraction algorithm.
No positional corrections for different colors of the stars were made. 

The following basic plate model with 8 parameters was used.

\[ \begin{array}{c@{\;=\;}*{8}{r}} 
   \xi  & a x & + b y & + c  & + e x & + f y & + p x^{2} & + q x y &   \\
   \eta &-b x & + a y & + d  & + f x & - e y &           & + p x y & + q y^{2} \\
   \end{array} \]

The 6 linear terms were split up into 4 orthogonal ($a,b,c,d$) and 
2 non--orthogonal terms ($e, f$).
The $p, q$ parameters determine the plate tilt.
The mean error of unit weight of the plate solutions was found
to be $\sigma_{0} = 199$ mas for the yellow and $203$ mas for the blue lens plates.

After the solution for the plate constants, observed places of
all stars were obtained, which then were transformed back
to mean places.

\subsection{Field Distortions}

The geometric distortions of the mapping process in the field of view
were corrected in 2 steps.
First, an approximate third--order optical distortion term was derived from
a pilot investigation, $D3 = -0.015 "/degree^{3}$ for the yellow and 
$-0.008 "/degree^{3}$ for the blue lens, and applied prior to the 
plate reduction with the basic model.
Second, the remaining field distortion pattern (FDP) was handled similar
to the investigation of the CPC2 data (\cite{CPC2FD}).
Over 50,000 residuals for each lens were averaged
as a function of the location on the plate.
The FDPs for both lenses are shown as vector plots in Fig.~4.
On the average about 300 stars contribute per vector.
Each diagram represents the sum of the effects as caused by one of the lenses
plus the measuring machine.
For subsequent reductions, these FDPs were subtracted from the $x,y$ data of
each individual plate.

\subsection{Magnitude Equations}

This is the most elaborate step in the data reduction.
Only preliminary corrections were applied for version 1.0 of the TAC.
Fig.~5 shows an example of a systematic error of the photographic positions
as a function of measured magnitude (original opacity values).
These magnitude equations have been found to depend at least on declination 
and epoch of observation.
The example in Fig.~5 shows all $\Delta \delta$ residuals
for the yellow lens plates taken between 1984.0 and  
1984.8 and with plate centers in the declination range 
$19.0^{\circ} \le \delta \le 46.0^{\circ}$.
Each dot represents the mean of 100 residuals.
From a least--squares fit with a linear model we obtained the parameters to
correct the x,y data.
The random errors are too large to show significant systematic errors for 
{\em individual} plates.
Thus, plates were empirically grouped by declination and epoch 
(see Fig.~1) for approximately constant systematic effect,
separately for both coordinates and both lenses.
About 30 groups per coordinate and lens were identified, containing
about 80 plates each.
A {\em single} linear correction model per coordinate was applied to 
{\em all} plates in each group. 
The $x$--coordinate of the yellow lens shows the largest magnitude equation.
Slopes up to 2.5 mas/opacity unit ($\approx 180$ mas/mag) were found 
with most values below 1.0. 

The plate adjustment was repeated after applying these corrections.
Due to the correlation with other parameters, the residual vs. measured magnitude
plots still showed a significant systematic error.
On purely empirical grounds, the originally obtained corrections were doubled
and applied to the original $x,y$ data. 
The magnitude equation plots after the re--reduction now showed nearly no residual
systematics.
The TAC positions, as obtained with the double corrections applied,
also gave better results than the uncorrected or straight corrected positions 
in external catalog comparisons.

\section{MEAN POSITIONS}

The mean error of a reference star position ($\sigma_{ref}$) 
is at least as large as the mean error of a $x,y$ measure ($\sigma_{xy}$). 
This leads to a significant error contribution, $\sigma_{pp}$, to the individual
photographic field star positions due to errors in the plate constants, $\sigma_{plc}$,
as obtained in the least--squares plate adjustment.  
Individual images close to the edge of the plate have a considerably larger
positional error than images close to the plate center due to the error
propagation law and the uncertainties in the derived plate constants.

Weights were calculated based on the quadratic sum of $\sigma_{xy}$ 
and $\sigma_{pp}$ as obtained from the error propagation law applied to the
basic plate model using $\sigma_{plc}$ for each individual plate and
the $x,y$ of each individual image.
Two catalogs were constructed initially, one from the plates of the
blue lens, the other from those of the yellow lens.
In each case the catalog position of a star is the weighted mean position of all
its images from different plates, excluding a few outlier observations.
In all cases (separate or combined catalog) the error of a TAC position,
$\sigma_{TAC}$, is given as the {\em standard error}
of the {\em mean position}, separately for each coordinate.

Catalog position differences (yellow $-$ blue) in 
$\Delta \alpha \cos \delta$ and $\Delta \delta$ 
were calculated for individual stars common to both catalogs.
There is still a systematic error as a function of magnitude for
$\Delta \delta$, as Fig.~6 shows, which varies with the 
{\em date of measurement}. 
No such significant error is found for the right ascension component.

Empirical corrections were applied to the $x,y$ data of all yellow
lens plates as a function of the date of measurement and catalog magnitude,
in order to bring those positions into the system of the blue lens plates.
Then all images were combined to form a single catalog, 
version 1.0 of the TAC,
using the same weighted mean procedure as explained above. 
The system of the blue lens plates was chosen because proper motions
of the TAC stars are derived by combining  with the AC,
which was observed in a blue spectral bandpass.
This "dirty" approach was undertaken in order to strengthen
the statistics for a single star and use on the average 4 to 5 images,
instead of having 2 separate catalogs with only about 2 images per
star and no handle on bad individual observations.
A future version of the TAC with better magnitude equation corrections
should give the same mean positions for the blue and yellow lens data.
The systematic offset found here is only due to insufficiently
accurate corrections applied for each lens separately.

\section{THE CATALOG}

The positions given in the TAC are for {\em equinox} J2000 and close to
the system of FK5 for the weighted mean {\em epoch of observation},
which may be different for each star.

This preliminary TAC with all measured data included,
was matched with individual zones of the re--reduced AC (\cite{urb95}).
   A search area with a radius of 14 arcsec was used to identify stars 
 from both catalogs.  If there is only one star from either catalog within the 
 search area, then the match is conclusive and can be used to compute a proper
 motion. If there is more than one star from either catalog within the search
 area, then the closest pair is used, provided that the separation is less than 
 half of the separation of the next closest match.  If the ratio of the separations 
 is greater than 0.5, then the match is ambiguous and a proper motion is not
 computed.
Graphical software was used to identify stars in complex multiple systems.

   Proper motions were determined as differences in position divided by the
 epoch differences. Errors to the proper motions, $\sigma_{pm}$,
 were calculated using 

\[ \sigma_{pm} \ = \ \frac{\sqrt{\sigma_{AC}^{2} + \sigma_{TAC}^{2}}}
			  {epoch_{TAC} - epoch_{AC}}  \]

where $\sigma_{TAC}$ is the mean error of the TAC position,
$\overline{\sigma_{TAC}}(mag)$, as a function of the
magnitude (see Table 4 below), scaled by the number of images, $n$,
for an individual star with respect to the average number of images
per star, which is 4.74,

\[  \sigma_{TAC} \ = \ \overline{\sigma_{TAC}} (mag) \ \sqrt{\frac{4.74}{n}} \]

and

\[ \sigma_{AC} \ = \ \frac{\overline{\sigma_{\alpha,\delta}}}
			  {\sqrt{n}}  \]

with $n$ being the number of images for that star in the AC and
$\overline{\sigma_{\alpha,\delta}}$ the average standard error per coordinate 
of an individual image in that zone as given in Table 2.
The precision of the AC positions is largely a function of the zone
and only to some extent dependent on magnitude.

This version of the TAC contains 705,679 stars including stars with
only 1 image and an average of 4.74 images per star. 
Fig.~7 shows the distribution of the standard errors in position of the
TAC stars, as obtained in this weighting mean procedure
of combining individual images of each star. 
The apparent overpopulation in the histograms around 170 mas is due to the
deliberately overestimated errors for those stars in the TAC with only 1 image.

The distribution of the number of images per star is given in Table 3.
The internal error of a TAC position, $\sigma_{TAC}$, at epoch of
observation as averaged over all stars is 85.9 mas in right ascension and
90.6 mas in declination. 
For the 591,648 stars of the TAC with 4 or more images per star
these numbers are 73.3 and 77.3 mas respectively.
On the average, the positions of the brighter stars are more precise than
those of the fainter stars.
Fig.~8 shows $\sigma_{TAC}$ for right ascension and declination as a
function of the catalog magnitude for {\em field stars} and the
number of stars per magnitude interval. 
These results are also summarized in Table 4.
The majority of stars are in the 10$-$11 magnitude range,
while the highest precision is at magnitudes 8$-$9.

Approximate magnitudes for individual images from each plate were
derived from the measured opacities using 
the Hipparcos Input Catalog (HIC, \cite{hic}) for calibration.
The blue and yellow bandpasses are very close to the Johnston B and V
photometric system.
This attempt should not be mistaken for a rigorous photometric reduction.
The TAC is an astrometric catalog and the approximate B and V
magnitudes are given mainly for the purpose of star identification
and a rough estimate of its color.

Table 5 summarizes some properties of the TAC.
The distribution of all TAC stars on the sky was plotted to
look for anomalies.
Fig.~9 shows an example for an $\approx 1900$ square degree area.
The band of the Milky Way and some star clusters are visible as well as 
the effects of different limiting magnitudes for different plates.
The average star density in the TAC is $\approx 30$ stars per
square degree.

TAC data for those stars which include proper motions derived from the AC
are available from the US Naval Observatory via anonymous ftp and the
World Wide Web (http://aries.usno.navy.mil/ad/tac.html) 
where an explanation of the data format is given as well.
More data will be released as the AC reductions progress.

\section{DISCUSSION}

\subsection{Random errors}

The standard error of unit weight of the plate reductions, $\sigma_{0}$,
is roughly the quadratic sum of $\sigma_{ref}$ and the mean random error of
the plate measuring process, $\sigma_{xy}$.
The reference star catalog (IRS) has an estimated positional error of
$\sigma_{ref} = 175$ mas per coordinate for a mean epoch of 1982 (\cite{co78}).
Thus from $\sigma_{0} = 200$ mas we obtain $\sigma_{xy} = 100$ mas,
which is close to the expected measuring error.

The positional error of a {\em field} star, $\sigma_{phot}$ is roughly the sum of
$\sigma_{xy}$ and the contribution $\sigma_{pp}$ due to the error propagation
from the plate constants.
The standard errors of the plate constants are given in Table 6.
The middle row in each section gives the average from all plate adjustments. 
The smallest and largest values of each parameter found are listed as well.
As an average over all plates,
using the basic plate model and the error propagation law we arrive 
at $\sigma_{pp} = 85$ mas for a star close to the plate center and
$\sigma_{pp} = 168$ mas for a star close to the edge of the plate.
Adding $\sigma_{xy} \approx 100$ mas, we obtain $\sigma_{phot} = 131$ mas and 
$196$ mas as a theoretical prediction for the precision of a single observation 
per coordinate.
From the scatter of the positions of individual images around the mean position 
we can obtain an observed value for this quantity.
The mean error of a catalog position of all those stars with at
least 4 images is $\sigma_{cat} \approx 75$ mas with an average 
of 5.25 images per stars. This gives $\sigma_{phot} \approx \ 172$ mas
for an estimate of the precision of a single observation, in agreement with the
theoretical value mentioned above.

\subsection{Systematic errors}

Because of the construction principle of a conventional plate adjustment,
the TAC positions are in the system of the reference star catalog used,
i.e. the IRS. This is close to but not identical to the J2000/FK5 system.
The FK5 system itself has local systematic errors of up to about 200 mas,
see e.g. (\cite{wobbles}), which will also be present in this TAC version.

In addition, remaining systematic errors as a function of magnitude can still
be found in version 1.0 of the TAC.
For comparison, the version 1.0 TAC data were split up into 
yellow and blue lens catalogs. 
Only stars with at least 2 images were used for this comparison.
Fig.~10 shows the position differences (yellow $-$ blue) lens catalog 
plotted vs. catalog magnitude, declination and right ascension.
The systematic pattern visible in some diagrams is the result of an 
insufficient (or over--) correction of magnitude--dependent systematic 
errors due to a lack of better reference stars.
The average added noise is on the $\pm 50$ mas level with amplitudes
to $100$ mas in some areas of the sky.

There is also a small systematic error in the TAC positions as a function
of color (spectral type) of the stars.
The rough estimate of magnitudes from the yellow and blue lens surveys
has not been used for an improved correction of refraction as a function
of color index at this point.
By construction, the positions in the TAC are in the system of the
reference stars. This is true for the separate blue and yellow data,
as well as for the combined catalog.
Thus, stars with a spectral type (color) close to the mean
color of the reference stars in a given area should not show systematic
offsets as a function of color.
Extreme blue or red stars in the TAC are displaced (mainly along declination)
with respect to the true geometric position, as would be seen from outside
the Earth's atmosphere.
The effect scales roughly with the tangent of the zenith distance, $z$.
For $z = 45^{\circ}$ we estimate a maximal displacement of $\pm 50$ mas
for extreme spectral types (B, M) for the blue lens data (\cite{bh96}).
For the yellow lens data the effect is only $\pm 13$ mas.
Thus, for the combined TAC as presented here we estimate a maximal
systematic error in position as a function of color of
about $\pm 40$ mas, while for most stars this will be below 20 mas. 
This is small as compared to the magnitude--dependent systematic
errors and the local distortions of the FK5/IRS system itself.
A better correction as a function of color is planned for a future 
version of the TAC, which will be on the Hipparcos system.

\subsection{External comparison with other catalogs}

External catalog comparisons were made with the combined (blue and yellow)
TAC version 1.0 data but including only stars with at least 2 images per star.
This TAC was compared with the PPM and ACRS, both in the northern,
$+5^{\circ} \le \delta \le 90^{\circ}$, and the southern hemisphere,
$ 0^{\circ} \ge \delta \ge -16^{\circ}$.
This choice of declination ranges restricts the influence of the high
precision CPC2 catalog (\cite{CPC2I}, \cite{CPC2II}) onto the southern
hemisphere section.
The catalog comparisons were made at the epochs of the individual TAC stars
using the proper motions from the other catalogs.
Only the high precision Part I of the ACRS was used for the comparisons,
which contains fewer stars than the PPM.

Table 7 summarizes the external catalog comparison results.
Estimates for the PPM accuracy were obtained from the published introductions
to the northern (\cite{PPMN}) and southern (\cite{PPMS}) volumes, propagated
to the mean epoch of the TAC at 1982.5.
Estimates for the ACRS accuracy were obtained by adopting
an average error of 90 mas at the central epoch  
and the published proper motion errors
(\cite{ACRS}) for the northern hemisphere. 
The southern hemisphere positions for 1982 are dominated by the CPC2,
which is included in both the PPM and the ACRS. 
Thus, the average error adopted here for the southern ACRS was that
for the PPM south.
Using 

\[  \sigma_{cat-TAC}^{2} \ = \ \sigma_{cat}^{2} \ + \ \sigma_{TAC}^{2}  \]

we derive an estimate of the TAC accuracy from each example.
Consistently the external accuracy of the TAC is found to be about 110 mas
per coordinate from all comparisons except for the ACRS north.
This discrepancy was traced to residual systematic errors of the faint
stars depending on magnitude in the Yale Zone data used for the ACRS.
The error budget for the brighter stars agrees with the other results;
only the fainter stars, which are the majority of stars used in the 
TAC$-$ACRS comparison, show systematic offsets in position,
mainly in the areas of the Yale Zones.
The objective of the ACRS is to provide a reference star catalog for the
reduction of the AC, thus the ACRS does not include any AC data,
contrary to the PPM. As a consequence it is much more difficult to obtain
proper motions for the ACRS as well as good positions for a recent epoch
like that of the TAC.
An external determination of the accuracy of all 3 catalogs from 3 comparisons,
e.g. (PPM$-$TAC), (ACRS$-$TAC), (PPM$-$ACRS) is not possible because of the
high correlation between the PPM and ACRS, which use many observation
catalogs in common. 
The TAC is a new observed catalog and not correlated to the ACRS or PPM.

Because all catalogs compared here are close to the FK5 system,
the local systematic errors of the FK5 system itself have to be added to obtain
the accuracy of the TAC with respect to an error--free coordinate system.
Furthermore, deviations of the IRS from the FK5 system are present. 
For stars in the magnitude range 9 to 10, the comparison TAC (yellow$-$blue) 
gives an estimate of the precision of 
$\sigma_{\alpha}\cos\delta = 111$ mas$ / \sqrt{2} = 78$ mas and
$\sigma_{\delta} = 129$ mas $/ \sqrt{2} = 91$ mas (Table 7).
This indicates larger residual systematic errors for the declination 
than for the right ascension component.

Fig.~11 and Fig.~12 show position difference vs. magnitude for both coordinates
and catalogs investigated here w.r.t. the version 1.0 TAC.
The overall agreement between catalogs is good except for the
southern hemisphere right ascension differences.
Fig.~13 similarly shows position differences vs. declination. 
The PPM$-$TAC differences show larger systematic variations on small scales
than the ACRS$-$TAC differences.

Plots of catalog differences vs. right ascension do not show overall
systematic features, except for the PPM(south)$-$TAC right ascension
component as displayed in Fig.~14.
This seasonal feature is not found in the ACRS(south)$-$TAC differences.

\section{CONCLUSION}

In order to serve the community we have decided to publish this
preliminary catalog now, rather than to wait for the additional data.
This TAC version 1.0 is incomplete in sky coverage as well as
in the correction for systematic errors.
Nevertheless, it represents a considerable improvement over
existing, similar astrometric catalogs, such as the PPM and the ACRS.
The precision of TAC version 1.0 is almost a factor of 3 
better than the previously best similar catalogs in the northern hemisphere
at current epochs and has a density of stars about 3 times as large. 

A future version of the TAC will be based on the Hipparcos
reference frame, thus all local systematic errors will
become negligible.  
In addition, the Hipparcos catalog provides enough high precision
reference stars to solve for systematic errors as a function of magnitude
on a plate--by--plate basis, which will greatly improve the precision 
of the TAC. 
The accuracy of this future TAC version is expected to be below 60 mas.
Simulations (\cite{zac92}) show a further improvement by using
block adjustment techniques, but the step from the IRS/FK5 to
the Hipparcos catalog gains the most.
The Tycho catalog, which will become available in summer 1997,
is more accurate than the TAC for most stars and comparable
to the TAC for the fainter stars.
The TAC will allow a realistic external comparison of the Tycho data
and will provide a high--weight third epoch set of positions between 
the Tycho and AC epochs. 
This is of particular interest for high proper motion
stars which will be detectable on the short (10 years) baseline
TAC$-$Tycho, but otherwise would be lost in ambiguous star matches 
when only the Tycho$-$AC data are compared.
The TAC contains about 20\% more stars than the Tycho catalog 
in those areas covered by the TAC, and proper motions can be derived 
for almost all TAC stars using the AC data.

\acknowledgments

The authors wish to acknowledge  
R.S.~Harrington\footnote{deceased in 1993}
for overseeing most of the observing and measuring part of 
this project and T.E.~Corbin, S.E.~Urban and 
J.A.~Hughes\footnote{deceased in 1992}
for valuable discussions.
The equipment was maintained by W.L.~Dunn, M.E.~Germain, J.~Hershey and C.~Hollins. 
Observing was accomplished by numerous members of the USNO 
Astrometry Department. 
Plate measuring was carried out by contract personnel, and by USNO Astrometry
staff members.
The automatic plate measuring was made possible by F.S.~Gauss.

\clearpage

\newpage

\figcaption[zacharias.fig1.ps]
  {Distribution of plate epochs
   as a function of declination of the plate centers.
   The diagram shows the distribution for the yellow lens plates, 
   which is almost the same as that of the blue lens plates because 
   of the simultaneous observing mode.}

\figcaption[zacharias.fig2.ps]
  {Sky distribution of plates used for the catalog.
   Plate centers are marked on an equal--area Aitoff projection 
   by a dot for the blue and a circle for the yellow plates.}

\figcaption[zacharias.fig3.ps]
  {Distribution of plate measure date as a function of declination
   of the plate centers, shown here for the yellow lens plates.
   The diagram is similar for the blue plates.}

\figcaption{Field distortion pattern (FDP) for 
   a) the blue lens, b) the yellow lens.
   The scale of the residuals is increased by a factor of 10,000.
   The largest vector is about 200 mas = 2 $\mu m$.
   These masks have been derived from reference star residuals of the
   conventional plate adjustments of all plates after correcting for a
   global third--order distortion term.}

\figcaption[zacharias.fig5.ps]
  {Systematic errors in position as a function of magnitude. 
   As an example, residuals (photographic $-$ reference star position)
   in declination are plotted vs.
   measured magnitude (opacity) for blue lens plates in the declination range of
   $19^{\circ} \le \delta \le 46^{\circ}$ and for observing epochs between 1984.0
   and 1984.8. One dot represents the mean of 100 residuals.}

\figcaption
  {Residual systematic errors as a function of magnitude depending on the
   date of measurement. 
   Plotted are the (yellow $-$ blue) lens catalog positions vs. magnitude for
   a) measure date day 15 in 1991 until end of 1993,
   b) measure date day 326 in 1994 until end of 1996.
   These systematic differences have been taken out before the
   yellow and blue lens data has been combined for the TAC version 1.0.}

\figcaption
  {Distribution of standard errors in positions for all TAC version 1.0 stars.
   a) $\sigma_{\alpha} \cos \delta$,  b) $\sigma_{\delta}$.}

\figcaption[zacharias.fig8.ps]
  {The precision of the TAC field star positions as a function of the catalog magnitude.
   Dots are for the right ascension and asterisks for the declination component.
   The approximate number of stars per magnitude interval is given as well.}

\figcaption[zacharias.fig9.ps]
  {Distribution of TAC stars on the sky. This example shows the area
   between $4^{h}$ and $8^{h}$ right ascension and $15^{\circ}$ to
   $45^{\circ}$ declination, using an equal--area Aitoff projection.}

\figcaption
  {Position differences (yellow $-$ blue) lens catalog in both coordinates
   plotted vs. catalog magnitude (a,b), declination (c,d) and right ascension
   (e,f). One dot represents the mean of 200 stars.
   Only stars with at least 2 images have been used for this comparison.
   The systematic pattern visible in some diagrams is the result of an insufficient 
   correction of mainly magnitude--dependent systematic errors due to a lack of better 
   reference stars.}

\figcaption
  {External catalog comparison. Differences $\Delta \alpha \cos \delta$
   vs. magnitude for a) ACRS(north)$-$TAC, b) PPM(north)$-$TAC, 
   c) ACRS(south)$-$TAC, d) PPM(south)$-$TAC.}

\figcaption
  {Similar to the previous figure but for the $\Delta \delta$ differences.}

\figcaption
  {External catalog differences vs. declination.
   a) $\Delta \alpha \cos \delta$ for ACRS$-$TAC, 
   b) $\Delta \alpha \cos \delta$ for PPM$-$TAC, 
   c) $\Delta \delta$ for ACRS$-$TAC,
   d) $\Delta \delta$ for PPM$-$TAC. }

\figcaption[zacharias.fig14.ps]
  {External catalog differences in $\Delta \delta$ vs. right ascension
   for PPM(south)$-$TAC.}


\begin{thebibliography}{}
\bibitem[Bastian \& R{\"o}ser 1993]{PPMS} Bastian,U., \& R{\"o}ser,S. 1993,
  PPM Star Catalogue, Vol.3,4, Astronomisches Rechen--Institut, Heidelberg
\bibitem[Corbin 1978]{co78} Corbin,T.E. 1978, in Modern Astrometry,
  Proceedings of IAU Colloq. No. 48, Vienna, edited by
  F.V.Prochazka and R.H.Tucker, Kluwer, Dordrecht, p.515
\bibitem[Corbin 1995]{co95} Corbin,T.E. 1995, IRS catalog from World Wide Web page 
  \begin{verbatim} http://aries.usno.navy.mil/ad/star_cats_rec.html \end{verbatim}
\bibitem[Corbin \& Urban 1990]{ACRS} Corbin,T.E., \& Urban,S.E. 1990,
  in Inertial Coordinate System on the Sky, Proceedings of IAU Symp. 141,
  ed. J.H.Lieske and V.K.Abalakin, Kluwer, Dordrecht, p. 433
\bibitem[de Vegt et al. 1993]{CPC2I} de Vegt,C., Murray,C.A., Zacharias,N., Nicholson,W.,
   Penston,M.J., \& Clube,S.M.V. 1993, A\&AS, 97, 985
\bibitem[Dieckvoss et al. 1975]{AGK3} Dieckvoss,W., Kox,H., G{\"u}nther,A., \&
   Brosterhus,E., 1975, AGK3, Hamburg--Bergedorf
\bibitem[Douglass \& Harrington 1990]{DH90} Douglass,G.G., \& Harrington,R.S. 1990,
   AJ 100, 1712
\bibitem[Hindsley 1996]{bh96} Hindsley,R. 1996, private communication
\bibitem[Kent 1994]{SDSS} Kent,S.M. 1994, ApSS, 217, 27
\bibitem[Perryman \& Hassan 1989]{Hip} Perryman,M.C.A., \& Hassan,H. (eds.) 1989,
The Hipparcos Mission, European Space Agency publication SP-1111, 3 volumes, Paris 
\bibitem[R{\"o}ser \& Bastian 1991]{PPMN} R{\"o}ser,S., \& Bastian,U. 1991,
  PPM Star Catalogue, Vol.1,2, Astronomisches Rechen--Institut, Heidelberg
\bibitem[Routly 1983]{allsky} Routly,P.M. 1983, in Sky with Ocean joined,
  Proceedings of the Sesquicentennial Symposium of the US Naval Observatory,
  edited by S.J.Dick and L. E.Doggett, Washington D.C., p. 145
\bibitem[Turon et al. 1992]{hic} Turon,C., et al. 1992, The Hipparcos Input Catalogue,
   European Space Agency (ESA) SP-1136, Volume 1-7, ESTEC, Noordwijk
\bibitem[Urban \& Corbin 1996]{urb95} Urban,S.E. \& Corbin,T.E. 1996, A \& A, 305, 989, 
   \begin{verbatim} http://aries.usno.navy.mil/ad/ac.html \end{verbatim}
\bibitem[Urban et al. 1996]{urb96} Urban,S.E., Martin,S.C., Jackson,E.S. \&
   Corbin,T.E. 1996, A\&AS, in press
\bibitem[Zacharias \& Douglass 1995]{ZD95} Zacharias,M.I., \& Douglass, G.G. 1995,
  Bull.AAS, 27, 857
\bibitem[Zacharias 1987]{HBAPP} Zacharias,N. 1987, dissertation, Univ.of Hamburg 
\bibitem[Zacharias 1992]{zac92} Zacharias,N. 1992, A\&A 264, 296  
\bibitem[Zacharias 1995]{CPC2FD} Zacharias,N. 1995, AJ, 109, 1880
\bibitem[Zacharias et al. 1992]{CPC2II} Zacharias,N., de Vegt,C., Nicholson,W., 
        \& Penston,M.J. 1992, A\&A, 254, 397
\bibitem[Zacharias et al. 1995]{wobbles} Zacharias,N., de Vegt,C., Winter,L.,
	\& Johnston,K.J. 1995, AJ, 110, 3093
\end{thebibliography}
\end{document}